# Layer-dependent Raman spectroscopy and electronic applications of wide-bandgap 2D semiconductor β-ZrNCl

Huiyu Nong[1], Qinke Wu[1], Junyang Tan[1], Yujie Sun[1], Rongxu Zheng[1], Rongjie Zhang[1], Shilong Zhao[1], and Bilu Liu[1, *]

[1]Shenzhen Geim Graphene Center, Tsinghua−Berkeley Shenzhen Institute and Institute of Materials Research, Tsinghua Shenzhen International Graduate School, Tsinghua University, Shenzhen, 518055, P. R. China

Correspondence should be addressed to B.L. (bilu.liu@sz.tsinghua.edu.cn)

**Abstract**

In recent years, two-dimensional (2D) layered semiconductors have received much attention for their potential in next-generation electronics and optoelectronics. Wide-bandgap 2D semiconductors are especially important in blue and ultraviolet wavelength region, while there are very few 2D materials in this region. Here, monolayer β-type zirconium nitride chloride (β-ZrNCl) is isolated for the first time, which is an air-stable layered material with a bandgap of ~3.0 eV in bulk. Systematical investigation of layer-dependent Raman scattering of ZrNCl from monolayer, bilayer, to bulk reveals a blue shift of its out-of-plane $A_{1g}$ peak at ~189 cm$^{-1}$. Importantly, this $A_{1g}$ peak is absent in monolayer, suggesting that it is a fingerprint to quickly identify monolayer and for the thickness determination of 2D ZrNCl. The back-gate field-effect transistor based on few-layer ZrNCl shows a high on/off ratio of 10$^8$. These results suggest the potential of 2D β-ZrNCl for electronic applications.

**Introduction**

Since graphene was discovered in 2004[1], a variety of two dimensional (2D) materials such as transition metal dichalcogenides (TMDCs)[2], hexagonal boron nitride (h-BN)[3], and black phosphorus[4,5] have received much attention. 2D materials are promising candidates for electronic, optoelectronic, and energy applications[6-8] due to their novel properties. For example, the dangling-bond free surface of 2D materials can suppress carrier trapping and ensure high performance in electrical transport at atomically-thin thickness. This feature is important to overcome the short channel effect during device scaling down, promoting the applications of 2D materials for field-effect transistor (FET), thin film transistor, photodetector, and phototransistor.[9-12] In addition, the planar structure of 2D materials makes them highly flexible and stackable[13], and gives rise to a freedom to manipulate the properties of 2D materials when their layer number changes[14-18]. For example, the layer-dependent magnetic ordering of 2D $CrI_3$ provides a platform to study physics of 2D magnetism[14]. In addition, $MoS_2$ exhibits an indirect-to-direct bandgap transition when it changes from bulk or bilayer to monolayer, which is desired in photoluminescence and optoelectronics[18]. The blooming research interest in 2D materials has driven the exploitation of 2D material family. Note that most studies of 2D semiconductors focus on narrow bandgap ($E_g$ <2 eV) semiconductors. For example, bandgaps of most TMDCs are 1-2 eV[2] and black phosphorus has a thickness-dependent bandgap of 0.3-2.0 eV[16]. For applications like high-power electronics, as well as optoelectronics for emitting, sensing and communication in blue and ultraviolet range[19-22], wide bandgap ($E_g$ >2 eV) materials are usually needed. Therefore, it is of interest to explore 2D semiconductors with wide bandgaps.

Along this direction, some wide bandgap 2D materials such as $GeI_2$[23] with a bandgap of 2.5 eV and InTeI[24] with a bandgap of 2.4-2.7 eV, have been theoretically predicted. However, it is still a challenge to synthesize and obtain ultrathin samples of these materials. Experimentally, a 2D layered III-IV metal chalcogenide family material such as GaS[25] with a bandgap of 3.05 eV and $SnS_2$[26-28] with a bandgap of 2.2 eV, have been obtained by exfoliation and chemical vapor deposition, respectively. FET devices based on these materials have been fabricated, showing an on/off ratio of ~$10^4$ and a mobility of 2.5 $cm^2$ $V^{-1}$ $s^{-1}$[26,27]. Wide bandgap 2D semiconductors with decent electrical properties are of great need while it is still a challenge. A layered material family of transition metal nitride halide (refer as MNX, where M=Hf, Zr, Ti, X=Cl, Br, I) catches our interest. They contains two polymorphs: orthogonal α-phase and rhombohedral β-phase, and β-phase MNX is usually stable in air and acidic solution[29]. β-ZrNCl has a wide bandgap of ~3 eV for bulk[30] and 3.3 eV for thin nanosheet[31], Theoretical study shows that monolayer β-ZrNCl has a bandgap of 2.93 eV based on Heyd-Scuseria-Ernzerhof (HSE) hybrid functional, which is larger than few layer and bulk materials in calculation[32], and shows great potential for next–generation electronic applications. MNX intercalated with metal ions is superconducting.[33-35] For thin materials, Ye et al.[36] have obtained 20-nm-thick β-ZrNCl flakes by micromechanical exfoliation and studied its superconductivity. Nakagawa et al.[37] have observed BCS-BEC crossover in intercalated β-ZrNCl nanoflakes. In another work, Feng et al.[31] have exfoliated β-ZrNCl into few layers by liquid phase exfoliation and studied its photothermal property. So far, the layer-dependent properties of 2D ZrNCl, especially down to monolayer, are still unclear.

In this study, we isolate monolayer and few layer β-ZrNCl by mechanical exfoliation, and study its layer-dependent Raman spectroscopy and electrical performance. The crystal structure of 2D ZrNCl is confirmed by X-ray diffraction (XRD), Raman spectroscopy, transition electron microscopy (TEM), scanning transition electron microscopy (STEM), energy-dispersive spectroscopy (EDS) and high-resolution atomic force microscopy (HRAFM). We observe a blue shift of ~3 cm$^{-1}$ for the out-of-plane $A_{1g}$ peak (189 cm$^{-1}$) of ZrNCl with increasing its thickness. Importantly, this $A_{1g}$ peak is absent in monolayer ZrNCl, which is a fingerprint for thickness identification. Furthermore, few-layer ZrNCl based FETs show good electrical properties with a high on/off ratio of $10^8$, showing its potential for electronic applications.

**Results and discussion**

The β-ZrNCl exhibits a rhombohedral SmSI crystal structure type (Figure 1a). One unit layer of ZrNCl is a sandwich structure made of Cl-N-Zr-Zr-N-Cl and can be viewed as nitride intercalation derivatives of ZrCl.[29] Each layer of ZrNCl is terminated by Cl atoms and interlayers connected by van der Waals force with a layer spacing of 0.92 nm. We first synthesized bulk ZrNCl crystals (Figure S1, Supporting Information) by a vapor transport method and its XRD pattern in the Figure 1b (the black curve) shows peaks at 9.59° (003), 19.30° (006), 29.02° (009), 31.43° (104), 39.01° (0012), and 50.60° (110). These peaks can be assigned to β-phase ZrNCl and is consistent with a previous study[38]. Figure 1c shows an optical microscopy image of the exfoliated few-layer ZrNCl flakes on SiO$_2$/Si substrate with a lateral size of ~20 μm. XRD pattern of the exfoliated 2D ZrNCl flakes (Figure 1b, the green curve) only exhibits {001} features, indicating the exfoliation is layer-by-layer and the (001) plane of the flakes are laying on the substrate. Good crystalline quality of 2D ZrNCl is confirmed by selective area electron diffraction and high-resolution STEM in Figure 1d, 1e, and Figure S2. The hexagonal symmetry is clearly shown in Figure 1d. In addition, the spacings of (110) plane and (210) plane are measured to be 0.18 nm and 0.32 nm, which agree with the XRD result[29]. EDS mapping is utilized to analyze the elements of Zr, N, and Cl, indicating a high uniform distribution of three elements in ZrNCl flakes (Figure S3). To directly witness the surface of as-exfoliated ZrNCl flakes, we perform HRAFM characterization (Figure 1f-h). The HRAFM image, the Fast-Fourier transformation (FFT) pattern, and the FFT-filtered HRAFM image together show hexagonal pattern with a spacing of 0.32 nm for (210) plane of ZrNCl flakes, which is consistent with STEM in Figure 1e. The lattice constant is calculated to be a=0.37 nm for ZrNCl by both STEM and HRAFM, which is consistent with XRD results. Taken together, these characterization results confirm the high crystallinity and clean surface of as-obtained 2D ZrNCl flakes.

Raman spectroscopy is a fast and non-destructive method to study layer-dependent properties of 2D materials and to identify their layer numbers. Here, we study the layer-dependent Raman scattering of monolayer, few-layer, and bulk ZrNCl. Figure 2a shows the optical microscope image of several ZrNCl flakes on $SiO_2$/Si substrates with different thicknesses, as evidenced by their colour. The thicknesses of the flakes are further measured by AFM along the yellow dash line in Figure 2b. We find that the thickness of monolayer ZrNCl is around 1.0 nm, which is close to the theoretical thickness of 0.92 nm in one unit layer[29]. The thicknesses of 2L and 4L ZrNCl flakes are 2.0 nm and 3.7 nm, respectively.

Figure 2c presents the Raman spectra of 1-4L and bulk ZrNCl excited by a 532 nm laser. For bulk sample, we observe 6 clear Raman active modes, namely the doubly degenerate in-plane $E_g$ modes at 126 cm$^{-1}$, 182 cm$^{-1}$, 590 cm$^{-1}$ and the out-of-plane $A_{1g}$ modes at 189 cm$^{-1}$, 331 cm$^{-1}$, 601 cm$^{-1}$, which is in agreement with the previous study[39]. For monolayer and few-layer sample, the $A_{1g}$ peak at 331 cm$^{-1}$ is too weak to be observed. Importantly, we find that the $A_{1g}$ peak at 189 cm$^{-1}$ is strong and sensitive to the thickness of ZrNCl (Figure 2c). Raman intensity mapping of this $A_{1g}$ peak (189 cm$^{-1}$) clearly shows its intensity increases with increasing flake thickness (Figure 2d). In addition, its peak position shows a blue shift of about 3 cm$^{-1}$ from 2L to bulk ZrNCl. According to the previous reports[31,39], $A_{1g}$ peak (189 cm$^{-1}$) comes from the out-of-plane vibration of outermost Cl atoms and it is obviously hardening due to the shortening of Zr-Cl distance paralleled to c axis with increasing thickness[31], consistent with our experiments. Furthermore, we find that the $A_{1g}$ peak (189 cm$^{-1}$) is absent in monolayer ZrNCl, which can be used as a fingerprint to quickly identify monolayer flakes. The above results indicate that Raman spectroscopy is a reliable tool to identify layer numbers of 2D ZrNCl.

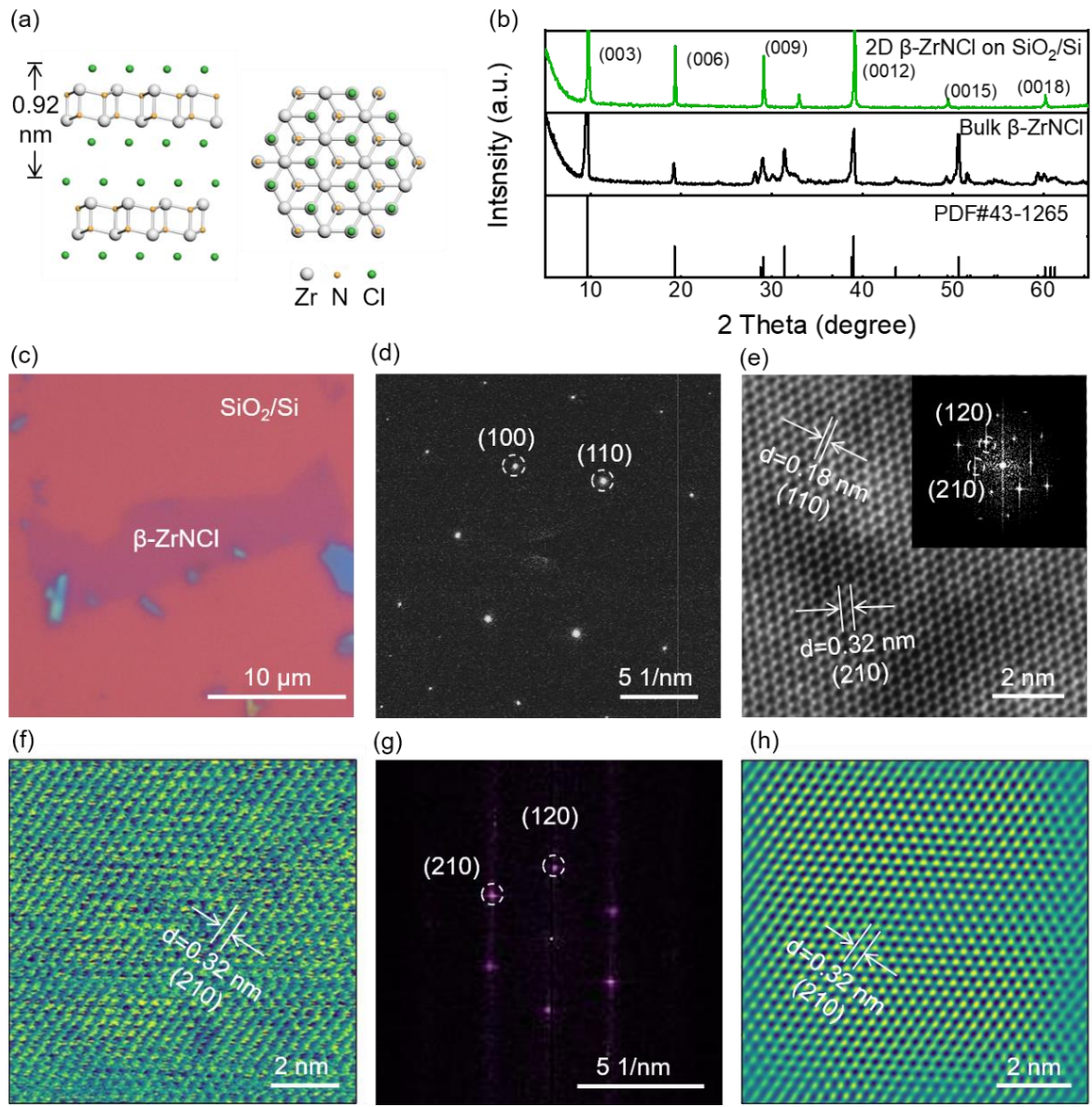

**Figure 1. Structure, exfoliation, and characterization of 2D β-ZrNCl.** (a) Structure of ZrNCl (left: side view, right: top view). (b) XRD patterns of 2D (green curve) and bulk (black curve) ZrNCl. (c) Optical microscopy image of the exfoliated ZrNCl flakes on SiO$_2$/Si substrate. (d) Diffraction pattern of 2D ZrNCl. (e) FFT-filtered STEM image of a ZrNCl flake and the inset is the corresponding FFT pattern. (f) HRAFM image, (g) FFT pattern, and (h) FFT-filtered HRAFM image of 2D ZrNCl.

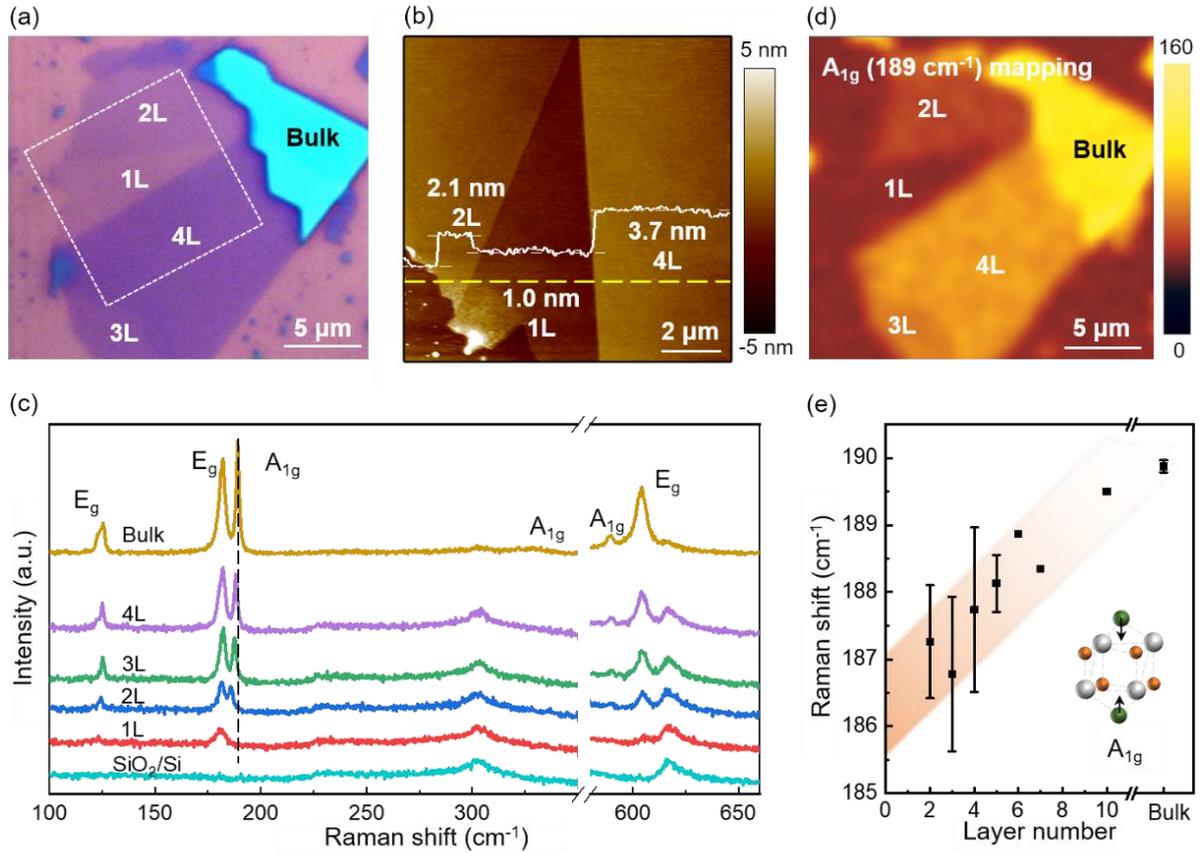

**Figure 2. Layer-dependent Raman spectroscopy of 2D β-ZrNCl.** (a) An optical microscope image of ZrNCl on SiO$_2$/Si substrate containing 1L, 2L, 3L, 4L flakes and bulk samples. (b) An AFM image of the white dotted area of (a) and the height profile taken along the yellow dash line. (c) Layer-dependent Raman spectra of ZrNCl from 1L, 2L, to bulk. Raman spectrum from bare SiO$_2$/Si substrate is also shown for comparison. (d) Raman intensity mapping of the A$_{1g}$ peak (~189 cm$^{-1}$) of the sample in (a). (e) A$_{1g}$ Peak position evolution as a function of layer numbers of ZrNCl.

Then, we focus on the environmental stability of 2D ZrNCl. We first prepare ZrNCl flakes and then use AFM and Raman spectroscopy to monitor the changes of their thickness, surface roughness, and crystal quality. Figure 3a-b are the optical microscope images of the ZrNCl flakes before and after exposure to air for 1 month. It is clear that the ZrNCl flakes still keep its integrity without noticeable damages or holes. AFM image of freshly exfoliated ZrNCl flakes (Figure 3c) and the flakes after air-exposure for 1 month (Figure 3d) show similar thickness and surface roughness (Figure 3e). Figure 3f shows the Raman spectra measured on the ZrNCl flakes before and after exposure to air for 3 months, showing similar features. These results indicate that 2D ZrNCl has a good stability in ambient condition, which is beneficial for practical applications.

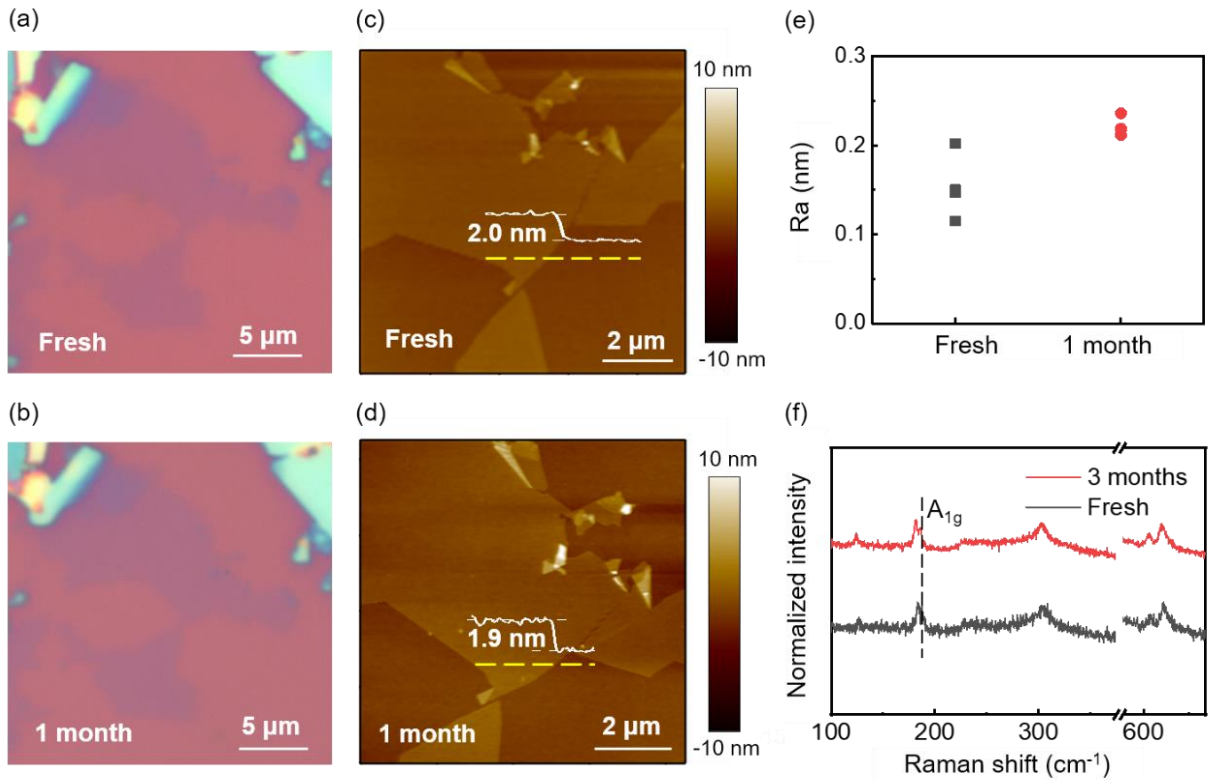

**Figure 3. Environmental stability of 2D β-ZrNCl.** (a) Optical microscope image and (c) AFM image of freshly exfoliated ZrNCl flakes. (c) Optical microscope image and (d) AFM image of the same flakes after exposure to ambient condition for 1 month. (e) Arithmetical mean roughness (Ra) of fresh ZrNCl and that after exposure to air for 1 month. Four 2×2 μm² areas of sample are calculated in (c) and (d), respectively. (f) Raman spectra of fresh ZrNCl flakes and that after exposure to air for 3 months. The intensity is normalized by the Si peak at 520.7 cm$^{-1}$.

Next, we fabricate back-gate FETs on the 2D ZrNCl to study its electrical properties. Figure 4a is the scheme of the FET device, and Figure 4b shows an optical microscope image of device based on a 3L ZrNCl flake. The electrodes are patterned by electron-beam lithography and thermally deposited with Cr/Au (10/30 nm). We use two electrodes of them (E1 and E2 in Figure 4b) to measure the electrical transport of the devices. Figure 4c shows the $I_d$-$V_g$ curves of three devices under bias voltage of $V_d$=1 V, and typical depleted n-type conduction behaviors are observed. The ZrNCl FETs show high on/off ratio of $10^8$, which is much higher than previous reported back-gate FET devices based on wide bandgap 2D materials, such as FL-GaS ($10^4$)[25], GaSe ($10^5$)[25] and $SnS_2$ ($10^4$)[26] (Figure 4d). The field effect mobility is calculated by $\mu=(dI_d/dV_g)\times L/(W\times C_g\times V_{ds})$, where L is the channel length, W is the channel width, $C_g$ is the gate capacitance of 285-nm-thick $SiO_2$ dielectric. The maximum electron mobility of 3.15 cm$^2$ V$^{-1}$ s$^{-1}$ is obtained, which is still much lower than theoretical predictions[32]. Figure 4e shows the typical $I_d$-$V_d$ curves of the device at gate voltages from -80 V to 80 V. The non-linear behavior indicates a Schottky contact formed at the interface of ZrNCl and metal contacts, where energy band of ZrNCl bends downward as illustrated in Figure 4f. These results indicate that the formation of Schottky barrier is responsible for the low mobility of the device. Further optimization of contacts and interface may improve device mobility of 2D ZrNCl.

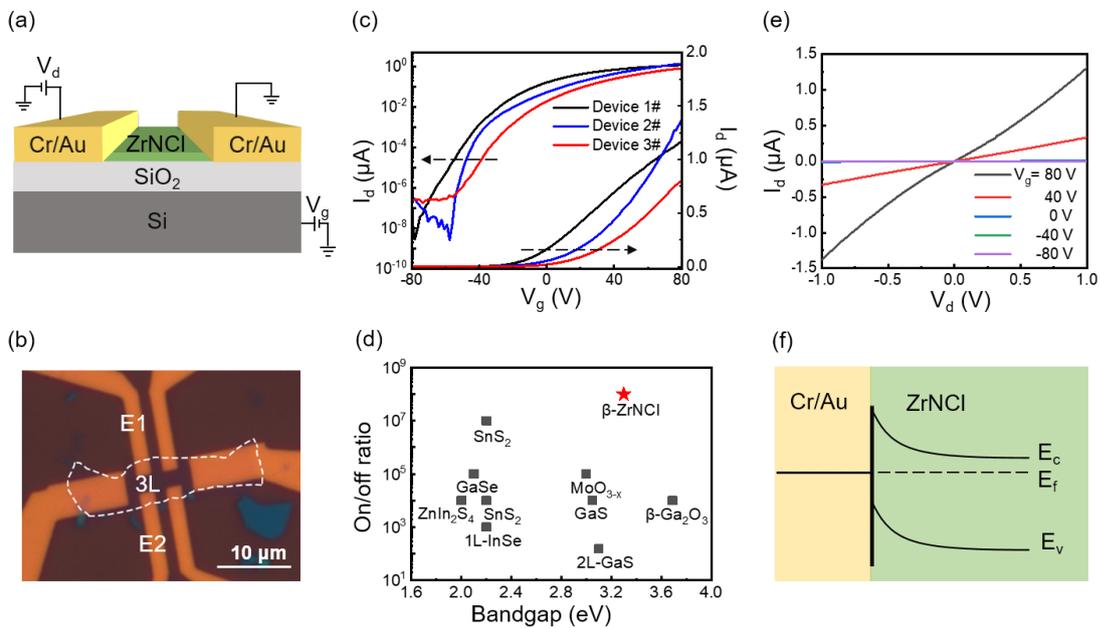

**Figure 4. Electrical transport measurements of few-layer β-ZrNCl back gate FET.** (a)(b) Schematic and optical microscope image of 3L ZrNCl FET. (c) $I_d$-$V_g$ curves of three different ZrNCl FETs in semi-log (left) and linear (right) scale. (d) A comparison of on/off ratios and bandgaps among 2D ZrNCl and the reported wide bandgap 2D materials[11,26,27,40-44]. (e) $I_d$-$V_d$ curves and (f) Band alignment diagram of ZrNCl FETs.

**Conclusions**

In summary, we have successfully isolated monolayer and few-layer β-ZrNCl and studied its layer-dependent properties for the first time. We find that the out-of-plane $A_{1g}$ peak at 189 cm$^{-1}$ shows an obvious blue shift by ~3 cm$^{-1}$ with increasing thickness of the ZrNCl flakes. Importantly, this $A_{1g}$ peak is absent in monolayer ZrNCl, suggesting it is a fingerprint to quickly identify monolayer samples and for thickness determination of ZrNCl. Meanwhile, few layer ZrNCl is found to be stable in ambient condition and back gate FETs based on ZrNCl shows a high on/off ratio of $10^8$. This work promotes the understanding of properties of ZrNCl at atomically thin thickness limit and shows its potential for future electronic and optoelectronic applications.

**Experimental Section**

*Synthesis of bulk β-ZrNCl crystals.* β-ZrNCl crystals were synthesized by a modified two-step method[45]. In the first step, Zr powders (~5 mg, Aladdin, 99.5%) were placed in a quartz boat at the center of a tube furnace. Then, NH$_4$Cl (200 mg, Aladdin, 99.5%) was placed in a quartz boat and put in the upstream of tube furnace. The furnace was heated to 700 °C within 20 min under Ar (100 sccm) and was kept for 15-30 min under Ar (100 sccm) and NH$_3$ (50 sccm). After reaction, NH$_3$ was turn off and the system was cooled down naturally under Ar. In the second step, product (40 mg) collected in the first step was mixed with NH$_4$Cl (20 mg, which was the transport agent), and was vacuum-sealed in a quartz ampoule with a diameter of 8 mm and a length of 40 mm. Then the ampoule was horizontally placed in a tube furnace and heated to 850 °C for 6 hours to grow ZrNCl crystals.

*Isolation and transfer of monolayer and few-layer ZrNCl.* The as-obtained ZrNCl crystals were exfoliated onto SiO$_2$/Si substrate using Scotch tape. For TEM characterization, 2D ZrNCl flakes on SiO$_2$/Si substrate were picked up by polyethylene terephthalate (PET) and were transferred onto TEM grid. Then PET was dissolved by soaking in dichloromethane (CH$_2$Cl$_2$).

*Materials characterization.* The morphology of the sample was characterized by optical microscope (Carl Zeiss Microscopy, Germany) and scanning electron microscope (Carl Zeiss Sigma 300, Germany). The thickness of sample was measured by AFM (Cypher ES, Asylum Research, USA). HRAFM image was obtained in a lateral force mode by using a soft tip (qp-BioAC, Nanosensors) with a spring constant of 0.06 N/m. The crystal structure of the sample was investigated by XRD (Bruker D8 Advance, with a monochromatic Cu Kα radiation λ = 0.15418 nm, Germany), TEM (FEI G2 spirit, 120 kV, USA), STEM (FEI Titan Themis G2 double aberration corrected TEM, 60 kV, USA). Raman spectra and mapping were collected by using a 532 nm laser excitation with a beam size of ~1 μm and a power of ~0.64 mW (Horiba LabRAB HR Evolution, Japan).

*Device fabrication and measurements.* Few-layer β-ZrNCl samples were exfoliated onto SiO$_2$/Si substrate with 285-nm-thick SiO$_2$ dielectric. Polymethyl methacrylate (PMMA) was spin-coated (8000 rpm for 1 min) onto the SiO$_2$/Si substrate with the samples as positive e-beam resist, and electron beam lithography was utilized to define the electrodes. Metal electrodes of Cr/Au (10/30 nm) were thermally deposited on the samples. Electrical transport measurements were performed using a semiconductor property analyzer (4200-SCS, Keithley, USA) in a vacuum probe station ($10^{-5}$ mBar, Lakeshore, USA) at room temperature.

**Acknowledgements**

We thank Prof. Andre Geim and Prof. Hui-Ming Cheng for discussions. The authors acknowledge the supports by the National Science Fund for Distinguished Young Scholars (No. 52125309), the National Natural Science Foundation of China (Nos. 51920105002, 51991343, and 51991340), Guangdong Innovative and Entrepreneurial Research Team Program (Program No. 2017ZT07C341), the Bureau of Industry and Information Technology of Shenzhen for the "2017 Graphene Manufacturing Innovation Center Project" (Project No. 201901171523), and the Shenzhen Basic Research Project (Nos. JCYJ20200109144616617, WDZC20200819095319002 and JCYJ20190809180605522)

**Supporting information**

Supporting Information is available from the Wiley Online Library or from the author.

**Conflict of interest**

The authors declare no conflict of interest.